\documentstyle[12pt]{article}
\newcommand{\beq}{\begin{equation}}
\newcommand{\eeq}{\end{equation}}
\newcommand{\bra}{\begin{array}}
\newcommand{\era}{\end{array}}

\title{ New algebraic structures in the $C_{\lambda}$-extended Hamiltonian system }
\author{E. H El Kinani$^{1}$}
\begin{document}
\date{}
\maketitle
\begin {center} {The Abdus Salam International Center for Theoretical Physics,
 ICTP-Strada costera 11, 34100 Tieste  Italy, \\ and \\ $^{(*)}$ Groupe
 de Math\'ematique Physique,  D\'epartement de Math\'ematiques, Facult\'e des
Sciences et Technique, Boutalamine B.P 509,  Errachidia, Morocco,\\

UFR, Physique th\'eorique, Facult\'e des Sciences, B.P. 1014,
Agdal, Rabat, Morocco.}
\end {center}
\vskip 1cm

{\bf {Abstract}}.\\

A realization of various algebraic structures in terms of  the
$C_{\lambda}$-extended oscillator algebras is introduced. In
particular, the $C_{\lambda}$-extended oscillator algebras
realization of Fairlie-Fletcher-Zachos (FFZ)algebra is given. This
latter lead easily to the realization of the quantum $U_t(sl(2))$
algebra.
The new deformed Virasoro algebra is also presented. \\
\vskip 6cm

\hrule
$^{(1)}$E-mail: hkinani@ictp.trieste.it or el-kinani@fste.ac.ma\\
$^{(*)}$Permanent address.

\newpage
\section {Introduction}
\hspace{.3in} Deformations of different groups and algebras have
attracted great attention  during the last few years. These new
mathematical object called quantum algebras or quantum groups have
found a lot of interesting physical applications. On the other
hand, recently various extensions and deformations of the
oscillator algebra have indeed been applied in the description of
systems with nonstandard statistics, with violation of the Pauli
principal, in the construction of integrable lattice models, as
well as the algebraic treatment of n-particle integrable models.
Among
the various deformations and extensions, we mention the following :\\

(i) The generalized deformed oscillator algebras {\bf
{(GDOA's)}}$^{\cite{1, 2, 3}}$, generated by the unit, creation,
annihilation, and number operators $(I, a, a^{+}, N)$ satisfying
the Hermiticity conditions:  $( a^{\dagger})^{\dagger}=a, N^{+}=N
$ and the commutation relations :

\begin{equation}
[\ N, a^{\pm}]=\pm a^{\pm}, \:\:\:\,\:\:\:\  [a,
a^{+}]_{q}=aa^{+}-qa^{+}a =F(N)
\end{equation}

 where $q$ is some real number and F(N) is some Hermitian, analytic function.\\

(ii) The G-extended oscillator algebras, where  G is some finite
group, appeared in connection with n-particle integrable systems.
For example, in the case of Calogero model $^{\cite{4, 5, 6}}$ G
is the symmetric group $S_{n}$. For two particles $S_{2}$ is
nothing but the cyclic group of order 2;  $C_{2}= \{ I, K, /K^{2}
=I \}$ and the obtaining  $S_{2}$-extended oscillator algebra  is
generated by the operators $(I, a, a^{+}, N$  and $K$) subject to
the Hermiticity conditions
     $( a^{+})^{+}=a , N^{+}=N $ and  $K^{+}=K ^{-}$ and the relations :

\begin{eqnarray}
\left[ N, a^{+}\right]=a^{+}, \:\:\:\ \left[N,K \right]=0, \:\:\ {K}^{2}=I  \\
\left[ a, a^{+} \right] = I+ r K  \:\:\ , \:\:\   {\mbox (r \in R)
}  \:\:\ a^{+}K= -K a^{+}
\end{eqnarray}

together with their Hermitian conjugates.\\

In this situation the Abelian group $S_{2}$ can be realized in
terms of Klein operator $K=(-1)^{N}$, where $N$ denotes the number
operators. Hence the $S_{2}$-extended oscillator algebra  becomes
a generalized deformed algebras, where $ F(N)=I+ r(-1)^{N}$ and
$q=1$ and known as the Calogero-Vasiliev or also modified
oscillator algebra $^{\cite{7,8}}$. By replacing $C_{2}$ by the
cyclic group of order $\lambda$ i.e $C_{\lambda}=\{ I, K,
K^{2},..., K^{\lambda-1} \}$, one get a new class of G-extended
oscillator algebras, generalizing the one describing the
two-particle Calogero-model. \\
The Letter is organized as follows
: in section 2 we review some basic notions concerning the
$C_{\lambda}$-extended oscillator algebras. Section 3 is devoted
to the construction of the FFZ and quantum $U_t(sl(2))$ algebra.
We propose a new deformed Virasoro algebras in Section 4.
Concluding remarks are given in the last section.
\section {Review on the $C_{\lambda}$-extended oscillator algebras and
its properties }

\hspace{.3in}In this section, we briefly review the relevant
definitions and results regarding the $C_{\lambda}$-extended
oscillator algebras ( for more details see ref. 9 and the
references quoted therein). The $C_{\lambda}$-extended oscillator
algebras $A^{\lambda}$, where $\lambda$ take any value in the set
$ \{2,3,...\}$, is defined as an algebra generated by the
operators $(I, a, a^{+}, N$ and $K$) subject to the Hermiticity
conditions $( a^{+})^{+}=a , N^{+}=N $ and  $K^{+}=K ^{-}$ and the
relations :

\begin{eqnarray}
\left[ N, a^{+}\right]=a^{+}, \:\:\:\ \left[N,K \right]=0, \:\:\ {K}^{\lambda}=I  \\
\left[ a, a^{+} \right] = I+ \sum_{r=1}^{\lambda-1}{\gamma}_{r}{K}^{r}  \:\:\ , \:\:\
 a^{+}K= e^{-2 \pi i / \lambda} K a^{+}
\end{eqnarray}

together with their Hermitian conjugates, where ${\gamma}_{r}$ are some complex
parameters restricted by the condition ${\gamma}_{r}^{*}={\gamma}_{\lambda-r}$,
and $K$ is the generator of cyclic group $C_{\lambda}$. For $\lambda=2$ we obtain
the Calogero algebra characterized by the commutation relations (2), (3).\\

Now let us examine the connection between the
$C_{\lambda}$-extended oscillator algebras and the generalized
deformed algebras {\bf {(GDOA's)}}. To begin, note that the cyclic
group that $C_{\lambda}$ has $\lambda $ inequivalente unitary
irreducible matrix representation $\Gamma^{\nu}$, ($\nu =0,1,2,..,
\lambda-1)$, which are one dimensional such that
$\Gamma^{\nu}({K^{r}})= exp(2\pi i \nu r / \lambda)$, for $r= 0,
1,2,...\lambda-1$. Hence the projection operator on the carrier
space of $\Gamma^{\nu}$ may be written as:

\begin{equation}
P_{\mu}={\frac{1}{\lambda}}\sum_{r=0}^{\lambda-1}(\Gamma^{\mu}({K^{r}}))^{*}
{K}^{r}= {\frac{1}{\lambda}}\sum_{r=0}^{\lambda-1} e^{-2 \pi i \mu r/{\lambda}} {K^{r}}
\end {equation}

and conversely we have :

\begin{equation}
 K^{r}= \sum_{\mu=0}^{\lambda-1} e^{2i \pi r \mu/{\lambda}}P_{\mu}.
\end {equation}

Then the algebra $A^{\lambda}$ equations (Eqs(4),(5)) can be
rewritten in terms of
 $ {I}, a, a^{+}, N $ and $P_{\mu}= P_{\mu}^{+}$ as follows :

\begin{eqnarray}
\left[ N, a^{+}\right]=a^{+}, \:\:\:\ \left[N,P_{\mu} \right]=0, \:\:\
\sum_{\mu=1}^{\lambda-1} P_{\mu}=I  \\
\left[ a, a^{+} \right] = I+ \sum_{\mu=1}^{\lambda-1}{\alpha}_{\mu}P_{\mu} \:\:\ , \:\:\
  \:\:\  a^{+}P_{\mu}= P_{\mu+1} a^{+} {\mbox and} \:\:\:\ P_{\mu}P_{\nu}= \delta_{\mu \nu}P_{\nu}
\end{eqnarray}

with the conventions $B_{\lambda}\sim B_{0}$ and $B_{-1} \sim
B_{\lambda -1}$ (where $B= P, \alpha$). The parameters $\alpha
_{\mu}$ are given by :

\begin{equation}
\alpha _{\mu}=\sum_{r=1}^{\lambda-1}exp(2i \pi  \mu r/{\lambda}){\gamma}_{r},
\end {equation}
restriced by the conditions $ \sum_{\mu=0}^{\lambda-1}{\alpha}_{\mu}=0$.
Hence we may eliminate one of them, for instance ${\alpha}_{\lambda-1}$.
In this situation the cyclic group generators $K$ and the projection
operators $P_{\mu }$ can be realized in terms of $N$ as : \\

\begin{equation}
K=e^{2 \pi iN / \lambda} \:\:\:\,\:\:\:\ P_{\mu}=\frac{1}{\lambda}
 \sum_{\nu=0}^{\lambda-1}e^{2 \pi i\nu (N-\mu) / \lambda} \:\:\, \:\:\
 (\mu, 1,2,...,\lambda-1)
\end {equation}
respectively. With such a choice, the algebra becomes a {\bf {(GDOA's)}}
where $G(N)= I+\sum_{\mu=0}^{\lambda-1}{\alpha}_{\mu}P_{\mu}$,
where $P_{\mu}$ is given by the above equation.\\

In the bosonic Fock space representation $^{\cite{9}}$,
we may consider the bosonic oscillator Hamiltonian, defined as usual by :

\begin{equation}
H_0 = \frac{1}{2}\{a^+ ,a \},
\end {equation}
which can be rewritten in terms of the projection operators as :

\begin{equation}
H_0 = N+ \frac{1}{2} I+ \sum_{\mu=0}^{\lambda-1}j_{\mu} P_{\mu}
\end {equation}
where $j_{0}=\frac{1}{2} \alpha_0$ and  $j_{\mu} =\sum_{\nu=0}^{\mu-1}
{\gamma}_{\nu}+ \frac{1}{2}{\gamma}_{\mu}$ for all $ \mu= 1,2,...\lambda-1$. \\

The eigenvectors  of $H_0$ are the states $|n>= |k \lambda +\mu>$
and the corresponding eingenvalues are given by :

\begin{equation}
E_{k \lambda +\mu}=k \lambda +\mu +j_{\mu}+\frac{1}{2}, \:\:\:\:\
k=0,1,..., \mu=0  \:\:\:\:\, 1,...\lambda-1
\end {equation}

In the each subspace of the $Z_\lambda$- graded Fock space, the
spectrum of $H_0$ is therefore harmonic, but the $\lambda$
infinite sets of equally spaced energy levels, corresponding to
$\mu =0,1,2,...\lambda-1$, may be shifted with respect to each
other by some amounts depending upon the algebra parameters
$j_{0}, ...j_{\lambda-2}$, through their linear combinations $
\alpha_{\mu}, \mu=0,1,...\lambda-1$.

In the case of Calogero-Vasiliev oscillator, the situation becomes
very simple and coincides with that of modified harmonic
oscillator.

\section {$C_\lambda$-extended oscillator realization of FFZ and $U_t(sl(2))$ symmetries.}
\subsection {$C_\lambda$-extended oscillator algebras realization of FFZ symmetry}

\hspace{.3in} Before going on, we would like to give a short
review concerning the $sdiff(X^{2n})$; algebra of volume
preserving diffeomorphisms on smooth manifold $X^{2n}$. Let
$X^{2n}$ be a $2n$-dimensional symplectic manifold with a
symplectic structure $\omega_{ab}$, which can be represented in
terms of the canonical constant antisymmetric $2n \times 2n$
matrix . Then $sdiff(X^{2n})$ is defined as

\begin{equation}
sdiff(X^{2n})=\{\Phi(\sigma)\in C^{\infty}(X)/[\Phi_1(\sigma),
\Phi_2(\sigma)]=w^{ab}\partial_{a}\Phi_{1} \partial_{b}\Phi_{2}
\},
\end{equation}
where $\sigma = (\sigma_{1},...,\sigma_{2n})$ denotes the
corresponding local coordinates on $X^{2n}$. In the simplest case
$X^{2n}= S^{1} \times S^{1}$, the Lie algebra elements of
$sdiff(T^{2})$ are given by \beq \Phi_n(\sigma)=\exp(\bf{n} \times
\sigma)\eeq where $\bf{n} \times
\sigma=n_{1}\sigma_{1}+n_{2}\sigma_{2}$, then $sdiff(T^{2})$ takes
the following form : \beq [\Phi_{\bf{m}}(\sigma),
\Phi_{\bf{n}}(\sigma)]=({\bf{m}} \Lambda
{\bf{n}})\Phi_{\bf{m}+\bf{n}}\eeq

This algebra has been studied first by Arnold $^{\cite{10}}$
 and investigated by other authors in the theory of relativistic
 surfaces( see  $^{\cite{11}}$ for more details).
The FFZ algebra or trigonometric sine algebra is defined as the
quantum deformation of the Lie algebra $sdiff(S^{1} \times
S^{1})$, which is generated by the generators $T_{\bf{m}}$
satisfying the following commutation relations

\begin{equation}
[T_{\bf{m}} ,T_{\bf{n}} ]=-2i\sin( \frac{2\pi}{\hbar}({\bf{m}}
\Lambda {\bf{n}})) T_{\bf{m}+\bf{n}} \:\:\:\:\, \hbar \in
{\mathcal{C}}^{*}
\end{equation}

One notes that the limit $\hbar \to 0$ reproduces the algebra (Eq.
(17)).Another approach to the definition of the above algebra is
based on the ideas of noncommutative geometry $^{\cite{12}}$.
Precisely, on the quantum two torus which is defined as an
associative $C^{*}$-algebra generated by two unitary generators
$U_1$ and $U_2$ satisfying the relations : $U_1 U_2=qU_2 U_1 $
where $(q=e^{i\hbar})$ .

Now we turn to discuss the $C_{\lambda}$-extended oscillator
realization of FFZ algebra. To begin with, let us define the
following operators:

\begin{equation}
 T_{\bf{m}}= e^{i \pi  m_{1} m_{2}/ \lambda }(a^+)^{m_1}(K)^{m_2}
\end{equation}
Before going on, let us discuss the problem of the negative power
of the reaction operators $a^{+}$, which makes this construction
formal. Indeed, to overcome this difficulty one consider the
Bargmann representation of the $C_{\lambda}$-extended oscillator
algebras given in $\cite{13}$. With such representation the
generators of $C_{\lambda}$-extended oscillator algebras are
identified with differential operators.\\
Using the relations (4), (5)one obtains:

\begin{equation}
 T_{\bf{m}} T_{\bf{n}}= e^{-i \pi  {\bf m} \Lambda {\bf n} / \lambda } T_{\bf{m}+\bf{m}}
\end{equation}

From the above equation, one easily gets the following relations

\begin{equation}
 [ T_{\bf{m}}, T_{\bf{n}}] = -2i \sin ({\frac{\pi}{\lambda}}
 ( {\bf{m}} \Lambda {\bf{n}} ))T_{\bf{m}+\bf{m}}
\end{equation}

So, the $T$'s satisfy the FFZ algebra (18), where we have
used the following change $2 \lambda=\hbar$. In what follows, we will
 generalize this construction for the $U_t(sl(2))$ algebra.\\

\subsection {$C_\lambda$-extended oscillator realization of $U_t(sl(2))$ symmetry}

\hspace{.3in} First let us  recall that the $U_t(sl(2))$ algebra
emerges in several contexts, e.g. in sine-Gordon theory $^{\cite
{14}}$ and in Chern-Simon theory $^{\cite{15}}$ and recently it is
uncovered in Landau problem which is intimately connected to
problem of fractional quantum Hall effect $^{\cite{16}}$. It is
well-known $^{\cite{15,17}}$ that the FFZ algebra induces the
quantum $U_t(sl(2))$ algebra. Relying on this fact,
 we present it $C_{\lambda}$-extended oscillator realization . To start,
 let's recall that the $U_t(sl(2))$ is defined as a complex unital associative
 algebra over $ {\bf{C}} (t)$, the field of fraction for the ring of formal
 power series in the indeterminate $t$ $(t \ne 0,1)$, generated by the generators
  $X^{\pm} , H$ and $H^{-1}$ satisfying : :\\

\begin{eqnarray}
H^{-1}H=HH^{-1}=1  \:\:\ &,&\:\:\ HX^{\pm}H^{-1}= t^{\pm 2}X^{\pm} \nonumber \\
\left[ X^{+},X^{-} \right]&=&{\frac{H-H^{-1}}{t-t^{-1}}}
\end{eqnarray}

Let us present the following construction depending on the pair
$(\bf{m},\bf{n})$ and the generators $T, X_{\pm}, H$ and $H^{-1}$

\begin{equation}
\begin{array}{ccc}
  X^+=\frac{1}{t-t^{-1}}(T_{\bf{m}}+T_{\bf{n}})
\\
&&\\

X^-=\frac{1}{t-t^{-1}}(T_{\bf{-m}}+ T_{\bf{-n}})
\\
&&\\

H=T_{\bf{m-n}}\:\:\:\:\ , \:\:\:\:\

H^{-1}=T_{\bf{n-m}}
\\
&&\\

\end{array}
\end{equation}

where the the deformation parameter $t= exp({-i \pi {\bf m}
\Lambda {\bf n}})/ \lambda$. Calculating the commutation relations
for $X^{\pm}$ and $H^{\pm}$ using  Eqs (19-21),
one get easily the commutation relation for $U_t(sl(2))$. \\

\section {$C_\lambda$-deformed Virasoro algebra.}

\hspace{.3in} In this section we introduce the new deformed
Virasoro algebra using  the $C_\lambda$-extended oscillator
algebras. To begin with, recall that the
 Virasoro algebra termed witt algebra or also conformal algebra was first introduced
  in the context of string theories. Its is relevant to any theory in 2-dimensional
  space-times which possesses conformal invariance.  The Witt algebra is the
  complexification of the Lie algebra $Vect(S^1)$. An element of $W$ is a linear
   combination of the elements of the form $e^{in \theta} \frac {d}{d \theta}$
   where $\theta $ is a real parameter and the Lie bracket on $W$ is given by :

\begin{equation}
 [e^{i m \theta} \frac {d}{d \theta}, e^{i n \theta} \frac {d}{d \theta}] =
 i(m-n)e^{i (n+m) \theta} \frac {d}{d \theta}
\end{equation}

Its is rather convenient to consider an embedding of the circle
into complex plane $C$ with the coordinates $z$, so that  $z=e^{i
\theta}$ and the element of the basis $e_m (m \in Z)$ are
expressed as $e_m = -z^{k+1}\partial_z$. In this basis the
commutation relations have the following form :
\begin{equation}
 [e_m, e_n] = (m-n)e_{m+n}
\end{equation}

On the other hand  the  deformation (q-deformation) of this
algebra was first introduced by Curtright and Zachos
$^{\cite{18}}$ and investigated in many occasion by many authors
 $^{\cite{19, 20, 21}}$, and defined by the following
 q-commutation relations:

\begin{equation}
 [e_m, e_n]_{q} =q^{m-n}e_m e_n-q^{n-m}e_n e_m= \frac {(q^{m-n}-q^{n-m})}{q - q^{-1}} e_{m+n}
\end{equation}

Turn now to the construction of  the $C_\lambda$-deformed Virasoro
algebra. To do this we will adopt the approach for undeformed case
$^{\cite{22}}$ where the generators $e_m$ are constructed from one
classical oscillator pair $a^{+}, a$ as the infinite-dimensional
extension of the following realization of $ sp(2) \sim o(2,1)$

\begin{equation}
 e_1=a, \:\:\:\ e_0=(a^+) a \:\:\:\ ,          e_1 =(a^+)^2 a
\end{equation}

The extension to positive indices $m$ is straightforward

\begin{equation}
 e_m = (a^+)^{m+1}a
\end{equation}

and the negative values $(m<-1)$ are  described by nonanalytic
dependence (monomials of $a^+$ with negatives powers). which acts
as the differential operators in Bargmann representation
$^{\cite{13}}$. From the following commutation relations between
the generators $a^{+}, a$ and $K$

\begin{equation}
\left[ a, a^{+} \right] = I+ \sum_{r=1}^{\lambda-1}{\gamma}_{r}{K}^{r} \:\:\ , \:\:\
 a^{+}K= e^{-2 \pi i / \lambda} K a^{+}
\end{equation}
one obtain after algebraic manipulation the following relations :

\begin{equation}
\left[ a, (a^{+})^{m^+1} \right] = (m +
\sum_{r=1}^{\lambda-1}f_{r}{\gamma}_{r} {K}^{r})(a^{+})^{m-1}
\end{equation}
where the $f_{r}$ is given by

\begin{equation}
f_{r}=(1+ e^{ r (2 \pi i / \lambda)}+ e^{ 2r(2 \pi i /
\lambda)}+...+e^{(m-1)r(2 \pi i / \lambda)}) \:\:\:\:\ f_{1}=1.
\end{equation}
Then from the previous equations, one get the following
commutation relation for generators $e_{m}(\gamma)$ :
\begin{equation}
 [e_m (\gamma), e_n (\gamma)] =(m-n) e_{m+n} (\gamma)+
 \sum_{r=1}^{\lambda-1}(e^{(n+1)r(2 \pi i / \lambda)}-e^{(m+1)r(2 \pi i
  / \lambda)}){\gamma}_{r} {K}^{r} e_{m+n} (\gamma).
\end{equation}

Which goes to the ordinary Virasoro algebra for ${\gamma}_{r} \to
0$. However, when ${\gamma}_{r} \ne 0$ it is something new, one
ask about the commutation relation between $K$ and the generators
$e_{m}$, thanks to Eqs.(5),(30) one easily finds :
\begin{equation}
\left[e_m (\gamma) , K \right] = g(\gamma)e_m (\gamma)K
\end{equation}

where $g(\gamma)=  (1-\exp 2\pi i(m+1)/{\lambda}) $. In the case
of  Calogero-Vasiliev, $\lambda=2$, in this situation the relation
(30) becomes :

\begin{equation}
\left[ a, (a^{+})^{m} \right] = (m+
\frac{1}{2}(1-(-1)^{m})){\gamma}_{1}{K})(a^{+})^{m-1},
\end{equation}
and the commutation relations between the  generators $e_m
(\gamma)$ are :

\begin{equation}
 [e_m (\gamma), e_n (\gamma)] =(m-n) e_{m+n}+ \frac{1}{2}((-1)^n -(-1)^m){\gamma}_{1}
 {K} e_{m+n} (\gamma).
\end{equation}

which is the $K$-deformed Virasoro algebra introduced in
$^{\cite{23}}$ . In this case, we have $g(2)=  (1+(-1)^m) $, hence
for $m$ odd, $e_{m}$ commutes with the operators $K$, for $m$ even
we have $g(2)=2$.

\section{Conclusion.}
In this letter, we have presented the realization of FFZ algebra
in terms $C_{\lambda}$-extended oscillator algebras, we have shown
how this realization lead to obtain the realization pf the quantum
quantum $U_t(sl(2))$. Otherwise,  we have presented the new
deformed Virasoro which we call the $C_{\lambda}$-deformed
Virasoro algebra. It is interesting to investigated these new
algebraic structures in the Calogero-Vasiliev model. Finally, note
that in same way one can
constructed $C_{\lambda}$-deformed $W$-algebras.\\

${\bf {Acknowledgments}}$ \\

The author would like to thank Prof. Randjbar-Daemi for his
invitation to the High Energy Physics Section of the Abdus Salam
Centre for theoretical Physics, Trieste, Italy where this work was
done.

\end{document}